\documentclass{appolb}
\usepackage{graphicx}
\usepackage{axodraw4j}
\usepackage{cite}
\newcommand{\bqa}{\begin{eqnarray}}
\newcommand{\eqa}{\end{eqnarray}}
\newcommand{\qbar}{\bar q}
\newcommand{\Eqn}[1]{Eq.~(\ref{#1})}
\newcommand{\Fign}[1]{Fig.~\ref{#1}}
\newcommand{\mur}{\mu_{\scriptscriptstyle R}}
\begin{document}
\title{Computing radiative corrections in four dimensions%
\thanks{Presented at the HiggsTools final meeting}%
}
\author{Roberto Pittau
\address{Departamento de F\'isica Te\'orica y del Cosmos and CAFPE, Universidad de Granada,
Campus Fuentenueva s.n., E-18071 Granada, Spain}
\\
}
\maketitle
\begin{abstract}
I comment and summarize the principles underlying the Four Dimensional Regularization/Renormalization (FDR) approach to the UV and IR infinities. A few recent results are also reviewed.
\end{abstract}
  
\section{Introduction}
The overwhelming complexity of the perturbative calculations performed nowadays to cope with the precision required by the present and future experimental measurements in High Energy Particle Physics makes it advisable to try alternative approaches to this problem. The source of many complications is the presence of divergent integrals in the intermediate steps of the calculations, that need to be regulated and removed from the physical predictions. 

In this contribution, I review the present status of the FDR approach~\cite{Pittau:2012zd}, with special emphasis on the mechanisms which should be used as guidelines when defining divergent integrals in unitary gauge theories. 
\section{FDR}
The main aim of FDR is embedding the UV subtraction directly in the definition of the loop integration. In that way, renormalized Green's functions are directly computed in four dimensions, without adding counterterms in the Lagrangian ${\cal L}$~\cite{Pittau:2013ica}. FDR can also be used to regulate IR divergences~\cite{Pittau:2013qla}. In the following two subsections, I review the main features of FDR. 
\subsection{UV infinities}
The FDR UV subtraction works at the integrand level. Consider, for example, 
a UV divergent one-loop integrand
\bqa
\label{eqn0}
J(q) = \frac{1}{q^2D_p}\,,~~~D_p = (q+p)^2.
\eqa
The FDR loop integration over $J(q)$ is defined as follows
\bqa
\label{eqn1}
\int [d^4q]\, J(q)
\equiv 
\lim_{\mu \to 0} \int_R
\left( J(q)-\frac{1}{\qbar^4}
\right)
\equiv 
\int [d^4q]\, \frac{1}{\qbar^2 \bar D_p}, 
\eqa 
where
\bqa
\qbar^2 \equiv q^2-\mu^2\,,~~~\bar D_p \equiv D_p-\mu^2.
\eqa
In \Eqn{eqn1} $R$ is an arbitrary UV regulator and $\mu^2$ regulates the IR behavior induced by the subtraction term $1/\qbar^4$.
Tensors are defined likewise. Given, for instance 
\bqa
J^{\alpha \beta}(q) = \frac{q^{\alpha}q^{\beta}}{q^2D_{p_1}D_{p_2}},
\eqa
one has
\bqa
\int [d^4q]\, J^{\alpha \beta}(q)
=
\lim_{\mu \to 0} \int_R
\left( J^{\alpha \beta}(q)-\frac{q^{\alpha}q^{\beta}}{\qbar^6}
\right).
\eqa 
This definition can be extended to more loops~\cite{Donati:2013voa}. The subtracted integrands are dubbed FDR vacua, or simply vacua, and do not depend on physical scales.
\subsection{Virtual IR divergences}
For IR convergent loop integrals, $q^2$ in the original integrands\,--\,such as $J(q)$ in \Eqn{eqn0}\,--\,can be left unbarred. Barring it regulates virtual IR divergences, giving rise to IR logarithms of $\mu$. As an example, the fully massless scalar one-loop triangle is defined in FDR as~\cite{Pittau:2013qla}
\bqa
\label{eq:cfun}
\int [d^4q]\frac{1}{\qbar^2 \bar D_{p_1} \bar D_{p_2}} 
\equiv \lim_{\mu \to 0} \int d^4q \frac{1}{\qbar^2 \bar D_{p_1} \bar D_{p_2}}=
\frac{i \pi^2}{2s} \ln^2\left(\frac{\mu^2}{-s-i0} \right),
\eqa
with $s= (p_1-p_2)^2= -2 (p_1\cdot p_2)$.
\subsection{Real IR divergences}
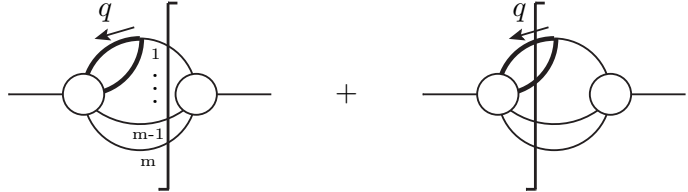
\begin{figure}[t]
\begin{center}
\begin{picture}(400,70)(0,20)
\SetScale{0.70}
\SetWidth{1}

\SetOffset(90,40) 
\SetWidth{1.5}
\Line(25,60)(30,60)
\Line(25,-40)(25,60)
\Line(20,-40)(25,-40)
\Text(11,23)[l]{\tiny 1}
\Text(11,15)[l]{\large .}
\Text(11,10)[l]{\large .}
\Text(11,5)[l]{\large .}
\Text(4,-7)[l]{\tiny m-1}
\Text(7,-18)[l]{\tiny m}
\SetWidth{1}
\Line(-60,11)(-20,11)
\SetWidth{2.5}
\Arc(-19,41)(30,270,360)
\Arc(10,11)(30,90,180)
\SetWidth{1}
\Arc(10,11)(30,0,90)
\Arc(10,11)(30,180,360)
\Line(40,11)(80,11)
\Text(80,8)[l]{\normalsize +}

\Text(-6,35)[b]{\normalsize $q$}
\LongArrow(7,47)(-10,42)

\Arc(10,35)(40,215,325)
\BCirc(-20,11){11}
\BCirc(40,11){11}

\SetOffset(245,40)

\SetWidth{1.5}
\Line(0,60)(5,60)
\Line(0,-40)(0,60)
\Line(-5,-40)(0,-40)
\SetWidth{1}
\Line(-60,11)(-20,11)
\SetWidth{2.5}
\Arc(-19,41)(30,270,360)
\Arc(10,11)(30,90,180)
\SetWidth{1}
\SetColor{Black}
\Arc(10,11)(30,0,90)
\SetColor{Black}
\SetWidth{1}
\Arc(10,11)(30,180,360)
\SetWidth{1}
\Line(40,11)(80,11)

\Text(-6,35)[b]{\normalsize $q$}
\LongArrow(7,47)(-10,42)

\Arc(10,35)(40,215,325)
\BCirc(-20,11){11}
\BCirc(40,11){11}
\end{picture}
\end{center}
\caption{\label{fig1} Cancellation of NLO final-state IR singularities in FDR.}
\end{figure}

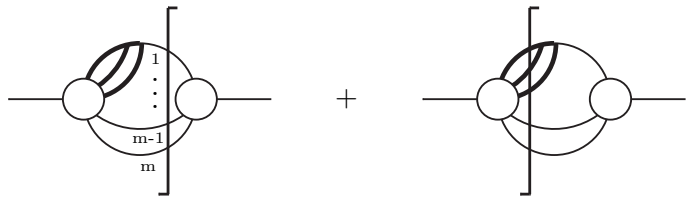
\begin{figure}[t]
\begin{center}
\begin{picture}(400,70)(0,20)
\SetScale{0.70}
\SetWidth{1}

\SetOffset(90,40) 
\SetWidth{1.5}
\Line(25,60)(30,60)
\Line(25,-40)(25,60)
\Line(20,-40)(25,-40)
\Text(11,23)[l]{\tiny 1}
\Text(11,15)[l]{\large .}
\Text(11,10)[l]{\large .}
\Text(11,5)[l]{\large .}
\Text(4,-7)[l]{\tiny m-1}
\Text(7,-18)[l]{\tiny m}
\SetWidth{1}
\Line(-60,11)(-20,11)
\SetWidth{2.5}
\Arc(-19,41)(30,270,360)
\Arc(10,11)(30,90,180)
\Arc(-35,51)(40,280,345)
%\Line(-20,11)(10,40)
%
\SetWidth{1}
\Arc(10,11)(30,0,90)
\Arc(10,11)(30,180,360)
\Line(40,11)(80,11)

%\LongArrow(7,47)(-10,42)

\Arc(10,35)(40,215,325)
\Text(80,8)[l]{\normalsize +}
\BCirc(-20,11){11}
\BCirc(40,11){11}
%\BCirc(10,40){5}

\SetOffset(245,40)

\SetWidth{1.5}
\Line(-3,60)(2,60)
\Line(-3,-40)(-3,60)
\Line(-8,-40)(-3,-40)
\SetWidth{1}
\Line(-60,11)(-20,11)
\SetWidth{2.5}
\Arc(-19,41)(30,270,360)
\Arc(10,11)(30,90,180)
\Arc(-35,51)(40,280,345)
%\Line(-20,11)(10,40)
%
\SetWidth{1}
\SetColor{Black}
\Arc(10,11)(30,0,90)
\SetColor{Black}
\SetWidth{1}
\Arc(10,11)(30,180,360)
\SetWidth{1}
\Line(40,11)(80,11)

%\LongArrow(7,47)(-10,42)

\Arc(10,35)(40,215,325)
\BCirc(-20,11){11}
\BCirc(40,11){11}
%\BCirc(10,40){5}
\end{picture}
\end{center}
\caption{\label{fig2} Cancellation of doubly unresolved final-state IR singularities in FDR (ansatz).}
\end{figure}

Virtual and real IR divergences are matched by a consistent treatment of the real radiation. The cutting rule
\bqa
\frac{i}{\qbar^2 + i0^+} \to (2\pi)\, \delta_+({\qbar^2})
\eqa
establishes the needed connection between barred loop propagators and massive external particles. 
As a consequence, the logarithms of $\mu$ in \Eqn{eq:cfun} can be rewritten as  
counterterms integrated over a $\mu$-massive phase-space $\bar \Phi_3$
\bqa
\label{eq:corr}
\int_{\Phi_2} \Re\left(\int [d^4q]\frac{1}{\qbar^2 \bar D_{p_1} \bar D_{p_2}} \right)
= \int_{\bar \Phi_3} \frac{1}{\bar s_{13} \bar s_{23}}~~~~ 
\left\{\hskip -5pt
\begin{tabular}{l}
$\bar s_{ij}= (\bar p_i+\bar p_j)^2$ \\
$\bar p^2_{i,j}=\mu^2$ 
\end{tabular}\right.. 
\eqa
Thus, {\Black $m$-}body virtual and {\Black $(m+1)$-}body real IR divergences compensate each other, as depicted in \Fign{fig1}. In both cases the divergent splitting is regulated by $\mu$-massive unobserved particles, denoted by thick lines.
This treatment has been shown to work at NLO~\cite{Pittau:2013qla}. The corresponding NNLO ansatz is illustrated in \Fign{fig2}.
   
\section{Fundamental properties of the loop integration}
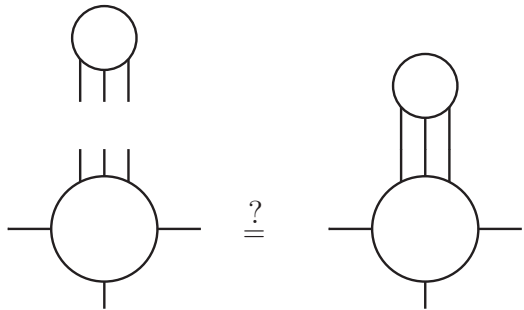
\begin{figure}[h]
\begin{center}
\begin{picture}(200,115) (0,-125)
\SetScale{0.6}
\SetWidth{1.5}
\SetOffset(40,-25)
\Line(15,0)(15,-40)
\Line(0,0)(0,-40)
\Line(-15,0)(-15,-40)
\BCirc(0,0){20}
\SetOffset(40,-97)
\Line(0,0)(60,0)
\Line(-60,0)(0,0)
\Line(15,0)(15,50)
\Line(0,0)(0,50)
\Line(-15,0)(-15,50)
\Line(0,0)(0,-50)
\BCirc(0,0){33}
%
% Right:
%
\SetOffset(160,-43)
\Line(15,0)(15,-40)
\Line(0,0)(0,-40)
\Line(-15,0)(-15,-40)
\BCirc(0,0){20}
\SetOffset(160,-97)
\Line(0,0)(60,0)
\Line(-60,0)(0,0)
\Line(15,0)(15,50)
\Line(0,0)(0,50)
\Line(-15,0)(-15,50)
\Line(0,0)(0,-50)
\BCirc(0,0){33}
\Text(-53,3)[r]{$~=^{\hskip -6.5pt \vphantom{A}^{\mbox{\large ?}}}~$} 
\end{picture}
\end{center}
\caption{\label{fig3} Schematic representation of the sub-integration 
consistency requirement.}
\end{figure}

In this section, I enumerate the three key properties that must be maintained by any consistent definition of loop integration and show how they are obeyed in Dimensional Regularization (DReg) and FDR. The properties are
\begin{enumerate}
\item Shift invariance;
\item The possibility of cancelling numerators and denominators;
\item The possibility of inserting sub-loop expressions in higher loop calculations (sub-integration consistency).     
\end{enumerate}
When the above requirements hold, rhs and lhs coincide in \Eqn{eqn2}, \Eqn{eqn3} and \Fign{fig3}, respectively.     
\bqa
\label{eqn2}
\int_R d^4q_1 \cdots 
d^4q_\ell\, 
J(q_1, \cdots,q_\ell)
=^{\hskip -6.5pt \vphantom{A}^{\mbox{\large ?}}}
\int_R d^4q_1 \cdots 
d^4q_\ell 
\,
J(q_1 + p_1, \cdots, q_\ell + p_\ell),
\eqa
\bqa
\label{eqn3}
\int_R d^4q_1 \cdots 
d^4q_\ell 
\frac{\rlap{\Red \bf \large/} D_i}{D_0\cdots {\rlap{\Red\bf \large /} D_i} \cdots D_k}
~=^{\hskip -6.5pt \vphantom{A}^{\mbox{\large ?}}}~
\int_R d^4q_1 \cdots 
d^4q_\ell 
\frac{1}{D_0\cdots D_k}, 
\eqa

The first condition guarantees routing invariance, the second one maintains the needed gauge cancellations, while the third requirement is essential to ensure unitarity. In fact, the unitarity equation
\bqa
T-T^\dag= i\, T^\dag T \nonumber
\eqa
mixes different loop orders, so that it is essential that the result of a sub-loop integration stays the same also when embedded in higher loop computations. 
\subsection{DReg}
In DReg the first two conditions are fulfilled by construction. On the other hand preserving the sub-integration consistency requires introducing order-by-order couterterms (CTs) in ${\cal L}$. For example without CTs one has
\bqa
\label{eqn4}
\left.\int d^nq_1 d^nq_2 \frac{1}{(q_1^2-M^2)^2} \frac{1}{(q_2^2-M^2)^2}\right|_{\frac{1}{\epsilon}=0}\!\!
\mbox{$\neq$}\, \left( \left.\int d^nq \frac{1}{(q^2-M^2)^2}\right|_{\frac{1}{\epsilon}=0}\right)^2,
\eqa
which prevents one from defining loop integrals as DReg integrals devoid of
$1/\epsilon$ poles.    
The role of the CTs is precisely subtracting UV poles is such a way to restore the equality in \Eqn{eqn4}.
\subsection{FDR}
FDR integrals are shift invariant, e.g.
\bqa
\int [d^4q]\, \frac{1}{\qbar^2 \bar D_p} = \int [d^4q]\, \frac{1}{\qbar^2 \bar D_{-p}}, 
\eqa
because both sides share the same subtraction term.
As for the numerator/denominator cancellation, one has to distinguish self-contractions of loop momenta generated by tensor decomposition from the case when they originate from Feynman rules. In the former case, no cancellation must occur.~\footnote{This is a consequence of requiring the result of the decomposition to coincide with the original tensor.} On the other hand, gauge invariance prescribes cancellation in the latter situation. FDR deals with both circumstances thanks to the introduction of the so called Extra Integrals (EI). Consider, for instance
\bqa
\int [d^4q] \frac{q^2}{\qbar^2 \bar D_{p_1} \bar D_{p_2}} \ne
\int [d^4q] \frac{\rlap{/} q^2}{\rlap{/} \qbar^2  \bar D_{p_1} \bar D_{p_2}}. 
\eqa
The inequality holds because the lhs subtracts $q^2/\qbar^6$, whilst $1/\qbar^4$ is subtracted in the rhs.
The difference can be computationally encoded in an EI, defined as the difference between the two subtraction terms surviving the $\mu \to 0$ limit
\bqa
\int [d^4q] \frac{\mu^2}{\qbar^2 \bar D_{p_1} \bar D_{p_2}} \equiv  \int_R d^4q 
\frac{\qbar^2-q^2}{\qbar^6}
=-\mu^2 \int d^4q \frac{1}{\qbar^6}=  \frac{i \pi^2}{2}.
\eqa
Thus, it is possible to write the following algebraic equation
\bqa
\int [d^4q] \frac{q^2}{\qbar^2 \bar D_{p_1} \bar D_{p_2}} =
\int [d^4q] \frac{\rlap{/} \qbar^2}{\rlap{/} \qbar^2  \bar D_{p_1} \bar D_{p_2}}+ 
\int [d^4q] \frac{\mu^2}{\qbar^2 \bar D_{p_1} \bar D_{p_2}}.
\eqa
From all of this, it is clear that preserving gauge cancellations prescribes
the replacement $q^2 \to \qbar^2$ both in denominators and numerators whenever 
$q^2$ does not originate from tensor reduction~\cite{Donati:2013iya}. This operation is called  Global Prescription (GP).  

As for the unitarity condition, the FDR counterpart of \Eqn{eqn4}
\bqa
\int [d^4q_1] [d^4q_2] \frac{1}{(\qbar_1^2-M^2)^2} \frac{1}{(\qbar_2^2-M^2)^2}
= \left(\int [d^4q] \frac{1}{(\qbar^2-M^2)^2}\right)^2,
\eqa
holds without the addition of CTs. However, the equality in
\Fign{fig3} is fulfilled only if the GP at the level of the sub-amplitude on the left does not clash with the GP at the level of the full amplitude on the right.
This is not always the case, but it is possible to correct for the mismatch and ensure sub-integration consistency by adding ``Extra''-Extra Integrals (EEI) derived by solely analyzing the loop diagrams on the right~\cite{Page:2015zca}.

\section{Results}
In the following, I review a few recent results 
obtained in the framework of FDR.
\subsection{DReg vs FDR @NLO}
A one-to-one correspondence exists between {DReg} and 
{FDR} for both UV and IR divergent loop {integrals}~\cite{Gnendiger:2017pys}
\bqa
{\Gamma(1-\epsilon)\, {\pi^\epsilon}} \int \frac{d^nq}{\mur^{-2 \epsilon}}~
\Big( \cdots \Big) \Bigg|_{\mur= \mu{\mbox{\,\,and $\frac{1}{\epsilon^i}$}=\,0}}= \int [d^4q]~ \Big( \cdots \Big).
\eqa
Analogously, for the real contribution
\bqa
\left(\frac{\mur^2}{s}\right)^\epsilon&& \int_{\phi_3} dx\, dy\, dz \Big( \cdots \Big) \delta(1-x-y-z) (xyz)^{-\epsilon}\Bigg|_{\mur= \mu{\mbox{\,\,and $\frac{1}{\epsilon^i}$}=\, 0}} \nonumber \\
= && \int_{\bar \phi_3} dx\, dy\, dz \Big( \cdots \Big) \delta(1-x-y-z+3 \mu^2/s), 
\eqa
where $\phi_3$ and $\bar \phi_3$ are massless and $\mu$-massive three-body phase spaces, respectively.
\subsection{DReg vs FDR @NNLO}
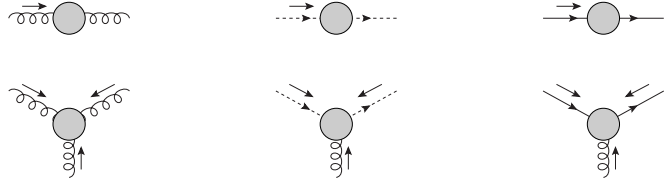
\begin{figure}[t]
\begin{center}
\begin{picture}(300,70)(-45,70)
\SetScale{0.5}
\SetWidth{0.5}
\SetOffset(-40,127)

\Gluon(45,0)(90,0){4}{4}
\Gluon(90,0)(135,0){4}{4}
\GCirc(90,0){12}{0.8}
%\Text(65,15)[b]{$k_1$}
\LongArrow(55,9)(70,9)
%\Text(90,-25)[t]{\large $G^{(\ell)}_1= G^{(\ell)}_{GG}$}
%
\SetOffset(60,127)
\DashArrowLine(45,0)(90,0){2.5}
\DashArrowLine(90,0)(135,0){2.5}
\GCirc(90,0){12}{0.8}
%\Text(65,15)[b]{$k_1$}
\LongArrow(55,9)(70,9)
%\Text(90,-25)[t]{\large $G^{(\ell)}_2= G^{(\ell)}_{cc}$}
%
\SetOffset(160,127)
\ArrowLine(45,0)(90,0)
\ArrowLine(90,0)(135,0)
\GCirc(90,0){12}{0.8}
%\Text(65,15)[b]{$k_1$}
\LongArrow(55,9)(70,9)
%\Text(90,-25)[t]{\large $G^{(\ell)}_3= G^{(\ell)}_{\Psi\Psi}$}
%
%
%
\SetOffset(-40,87)
\Gluon(45,25)(90,0){4}{4}
\Gluon(90,0)(135,25){4}{4}
\Gluon(90,0)(90,-40){4}{4}
\GCirc(90,0){12}{0.8}
%
%\Text(67,30)[b]{$k_1$}
\LongArrow(55,30)(70,22)
%\Text(110,30)[b]{$k_2$}
\LongArrow(124,30)(109,22)
%\Text(104,-28)[l]{$k_3$}
\LongArrow(99,-34)(99,-20)
%
%\Text(90,-53)[t]{\large $G^{(\ell)}_4= G^{(\ell)}_{GGG}$}
%
\SetOffset(60,87)
\DashArrowLine(45,25)(90,0){2.5}
\DashArrowLine(90,0)(135,25){2.5}
\Gluon(90,0)(90,-40){4}{4}
\GCirc(90,0){12}{0.8}%
%
%\Text(67,30)[b]{$k_1$}
\LongArrow(55,30)(70,22)
%\Text(110,30)[b]{$k_2$}
\LongArrow(124,30)(109,22)
%\Text(104,-28)[l]{$k_3$}
\LongArrow(99,-34)(99,-20)
%
%\Text(90,-53)[t]{\large $G^{(\ell)}_5= G^{(\ell)}_{Gcc}$}

%
\SetOffset(160,87)
\ArrowLine(45,25)(90,0)
\ArrowLine(90,0)(135,25)
\Gluon(90,0)(90,-40){4}{4}
\GCirc(90,0){12}{0.8}
%
%\Text(67,30)[b]{$k_1$}
\LongArrow(55,30)(70,22)
%\Text(110,30)[b]{$k_2$} 
\LongArrow(124,30)(109,22)
%\Text(104,-28)[l]{$k_3$}
\LongArrow(99,-34)(99,-20)
\end{picture}
\end{center}
\caption{\label{QCDcorr} Irreducible 2-  and 3-point QCD Green's functions}
\end{figure}

FDR has been proven to renormalize consistently off-shell QCD up to two loops~\cite{Page:2015zca}. The {$\alpha_S$} and {$m_q$} shifts necessary to translate
{FDR} to {$\overline{{\rm \Black MS}}$} in DReg have been determined by analyzing the FDR vacua of the two-loop 2- and 3-point QCD correlators $G^{({\rm 2-loop})}$ given in \Fign{QCDcorr}.

\subsection{EEIs}
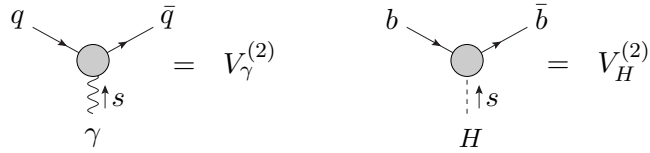
\begin{figure}[t]
\begin{center}
\begin{picture}(300,60)(-120,125)
\SetScale{0.5}
\SetWidth{0.5}
\SetOffset(-100,150)
\Text(17,12)[b]{$q$}
\Text(73,12)[b]{$\bar q$} 
\Text(42,-29)[l]{$\gamma$}
\Text(52,-15)[l]{$s$}
\Text(75,0)[l]{$=~~V^{(2)}_\gamma$}
\ArrowLine(45,25)(90,0)
\ArrowLine(90,0)(135,25)
%\Gluon(90,0)(90,-40){4}{4}
\Photon(90,0)(90,-40){4}{4}
\GCirc(90,0){12}{0.8}
%
%\Text(67,30)[b]{$k_1$}
%\LongArrow(55,30)(70,22)
%\Text(110,30)[b]{$k_2$} 
%\LongArrow(124,30)(109,22)
%\Text(104,-28)[l]{$k_3$}
\LongArrow(99,-36)(99,-22)
%
%\Text(90,-53)[t]{\large $G^{(\ell)}_6= G^{(\ell)}_{G\Psi\Psi}$}
%
%
%
\SetOffset(40,150)
\Text(17,12)[b]{$b$}
\Text(73,12)[b]{$\bar b$} 
\Text(42,-29)[l]{\small $H$}
\Text(52,-15)[l]{$s$}
\Text(75,0)[l]{$=~~V^{(2)}_{\small H}$}
\ArrowLine(45,25)(90,0)
\ArrowLine(90,0)(135,25)
%\Gluon(90,0)(90,-40){4}{4}
%\Photon(90,0)(90,-40){4}{4}
\DashLine(90,0)(90,-40){4}
\GCirc(90,0){12}{0.8}
%
%\Text(67,30)[b]{$k_1$}
%\LongArrow(55,30)(70,22)
%\Text(110,30)[b]{$k_2$} 
%\LongArrow(124,30)(109,22)
%\Text(104,-28)[l]{$k_3$}
\LongArrow(99,-36)(99,-22)
%
%\Text(90,-53)[t]{\large $G^{(\ell)}_6= G^{(\ell)}_{G\Psi\Psi}$}
\end{picture}
\end{center}
\caption{\label{fig4} On-shell two-loop $\gamma^\ast \to q \bar q$ and $H \to b \bar b$ QCD vertices.}

\end{figure}

Analyzing the FDR EEIs led to a fix of two-loop {\em ``naive''} FDH 
in DReg~\cite{Page:2015zca}:
\bqa
G^{({\rm 2-loop})}_{\rm{bare,\,DReg}}|_{n_s=4} 
\rightarrow 
G^{({\rm 2-loop})}_{\rm{bare,\,DReg}}|_{n_s=4} +   {\textstyle \sum_{\rm{Diag}}} 
{\rm EEI}_b|_{n_s=4},
\eqa
where $~n_s= \gamma_\mu \gamma^\mu= g_{\mu \nu}g^{\mu \nu}$.
In the above equation,  EEI$_b$s are DReg integrals obtained from FDR EEIs by 
dropping the subtraction term, e.g.
\bqa
\int [d^4q] 
\frac{1}{\qbar^2 \bar D_p} \to  \int d^nq
\frac{1}{q^2 D_p}.
\eqa
The EEI$_b$s reproduce the effect of the evanescent operators needed
in FDH and dimensional reduction to restore renormalizability, at least off shell. A preliminary study of the two-loop QCD vertices in \Fign{fig4}
indicates that the same phenomenon is likely to be observed
on-shell as well~\cite{private}.
\subsection{Local subtraction of IR divergences @NLO}
It is possible to set up a local FDR subtraction of the final-state IR infinities by rewriting the virtual logarithms as counterterms to be added to the real radiation~\cite{Gnendiger:2017pys}, in the same spirit of \Eqn{eq:corr}. Schematically  
\bqa
\label{eqn:snlo}
\sigma_{\rm{NLO} } &=& 
\int_{\Phi_2}
\Bigl(
|M|^2_{\rm{Born} } + 
\underbrace{|M|^2_{\rm{Virt}}}_{\rm{devoid~of~logs~of~\mu}}
\Bigr)  F_J^{\rm{\tiny(2)}}(p_1,p_2) \nonumber \\
&+& \!\!\!\!\!\!\!
\underbrace{\int_{\Phi_3}}_{\rm{\mu \to 0~here}} 
\Bigl(
 |M|^2_{\rm{Real}}\, F_J^{\rm{(3)}}(p_1,p_2,p_3) 
-|M|^2_{\rm{CT}}  
\,F_J^{\rm{(2)}}(\!\!\!\!\!\!\underbrace{\hat p_1,\hat p_2}_{\rm{mapped~kinematics}}\!\!\!\!\!\!)
\Bigr),
\eqa
where $F_J$ are jet functions.
For instance, in the case of $e^+ e^- \to \gamma^\ast \to q \bar q(g)$, the explicit form of the local counterterm is
\bqa
\label{eq:ct}
|M|^2_{\rm{CT}} &=& \frac{16 \pi \alpha_s}{s} C_F
|M|^2_{\rm{Born}}({\hat{p}_1,\hat{p}_2}) 
\Biggl(
\frac{s^2}{ s_{13}  s_{23}}
-\frac{s}{ s_{13}}
-\frac{s}{ s_{23}} \nonumber \\
&&+ \frac{ s_{13}}{2 s_{23}}
+ \frac{ s_{23}}{2 s_{13}}
-\frac{17}{2}
\Biggr),
\eqa
and the mapping reads
\bqa
\hat{p}^{\,\alpha}_1= \kappa \Lambda^{\!\alpha}_{\,\beta}\, p^\beta_1
\left(1+\frac{s_{23}}{s_{12}}\right),~~
\hat{p}^{\,\alpha}_2= \kappa \Lambda^{\!\alpha}_{\,\beta}\, p^\beta_2
\left(1+\frac{s_{13}}{s_{12}}\right),
\eqa
where $\kappa= \sqrt{\frac{s s_{12}}{(s_{12}+s_{13})(s_{12}+s_{23})}}$ and 
$\Lambda^{\!\alpha}_{\,\beta}$ is the boost that brings  the sum of 
$\hat{p}_1$ and $\hat{p}_2$ back to the center of mass frame:
$\hat{p}_1+\hat{p}_2= (\sqrt{s},0,0,0)$.

The inclusive $\sigma_{\rm  \mbox{\tiny NLO}}= \sigma_0 
\left(1+C_{\mbox{\tiny \rm F}} \frac{3}{4}
\frac{\alpha_s}{\pi}\right)$ cross section is reproduced by a numerical implementation of \Eqn{eqn:snlo}. In addition, successful comparisons~\cite{prep} with  
{\tt MadGraph5\_aMC@NLO}~\cite{Alwall:2014hca} interfaced with {\tt FastJet}~\cite{Cacciari:2011ma} have been attained for realistic jet observables.

\section{Outlook}
FDR is turning to a competitive tool to compute radiative corrections.
The UV subtraction is incorporated, at the integrand level, in the definition of the loop integration. As a consequence, one directly deals with four-dimensional integrals, without introducing UV counterterms in ${\cal L}$. This has been shown to be a workable alternative to DReg up to two loops for off-shell quantities. 
 
The FDR regularization of the IR divergences is well understood at NLO, and a completely local subtraction of final-state IR infinities has been worked out for two-jet cross sections.   

Going on-shell at NNLO seems feasible. In fact, as a by-product of the FDR UV treatment, a fix to two-loop {\em ``naive''} FDH avoiding evanescent couplings is available for on-shell observables.

Future investigations include an extension of FDR to initial-state IR singularities and a complete two-loop calculation~\cite{bprp} of the QCD form factors in \Fign{fig4}.
Finally, it would be interesting to investigate FDR integration as a new mathematical tool to be used also in other branches of Physics where divergent integrals occur.

\end{document}